\documentclass[fleqn,11pt]{article}

\usepackage{amsfonts,amsmath,amssymb}
\usepackage{latexsym}

\usepackage{amsfonts,amssymb,cite}
\usepackage{graphicx}

\def\noi{\noindent}
\def\jnumber#1#2{\thispagestyle{empty} \noi\unitlength=1mm
    	\begin{picture}(178,10)
            \put(177,15){\llap{\large\it Grav. Cosmol. No.\,#1, #2}}
                    \end{picture}}

\newcommand{\Title}[1]{\noi {{\Large\bf #1}}\\[1ex]}

\def\Aunames#1{\noi{\bf #1}}
\def\au#1{${}^{#1}$}
\def\Addresses#1{\medskip\noi \protect
	\begin{description}\itemsep -3pt {\it #1} \end{description}}
\def\adr#1#2{\item[${}^{#1}$]{\it #2}}

\newcommand{\Abstract}[1]{\vskip 2mm \begin{center}
        \parbox{16.4cm}{\small\noi #1} \end{center}\medskip}

\def\email#1#2{\footnotetext[#1]{e-mail: #2}\addtocounter{footnote}{1}}

\def\nqq{\hspace*{-2em}}

\def\cm{\hspace*{1cm}}

\usepackage{color}

\def\Acknow#1{\subsection*{Acknowledgments} #1}
\def\Funding#1{\subsection*{Funding} #1}

\def\Jl#1#2{#1 {\bf #2},\ }

\def\ApJ#1 {\Jl{Astroph. J.}{#1}}
\def\CQG#1 {\Jl{Class. Quantum Grav.}{#1}}
\def\DAN#1 {\Jl{Dokl. AN SSSR}{#1}}
\def\GC#1 {\Jl{Grav. Cosmol.}{#1}}
\def\GRG#1 {\Jl{Gen. Rel. Grav.}{#1}}
\def\IJMPD#1 {\Jl{Int. J. Mod. Phys. D}{#1}}
\def\JETF#1 {\Jl{Zh. Eksp. Teor. Fiz.}{#1}}
\def\JETP#1 {\Jl{Sov. Phys. JETP}{#1}}
\def\JHEP#1 {\Jl{JHEP}{#1}}
\def\JMP#1 {\Jl{J. Math. Phys.}{#1}}
\def\NPB#1 {\Jl{Nucl. Phys. B}{#1}}
\def\NP#1 {\Jl{Nucl. Phys.}{#1}}
\def\PLA#1 {\Jl{Phys. Lett. A}{#1}}
\def\PLB#1 {\Jl{Phys. Lett. B}{#1}}
\def\PRD#1 {\Jl{Phys. Rev. D}{#1}}
\def\PRL#1 {\Jl{Phys. Rev. Lett.}{#1}}

\def\lal{&&\nqq {}}

\def\beq{\begin{equation}}
\def\eeq{\end{equation}}
\def\bear{\begin{eqnarray}}
\def\bearr{\begin{eqnarray} \lal}
\def\ear{\end{eqnarray}}
\def\earn{\nonumber \end{eqnarray}}

\def\nnn{\nonumber\\ \lal }

\begin{document}
\twocolumn[
\jnumber{issue}{year}

\Title{Bondi accretion onto Damour-Solodukhin wormhole}   

\Aunames{R.~M. Yusupova,\au{a,b,1} R.~Kh. Karimov,\au{b,2} and A. Bhattacharya\au{c,3}}

\Addresses{
\adr a {Institute of Molecule and Crystal Physics, Ufa Federal Research Centre, Russian Academy of Sciences, Prospekt Oktyabrya 151, Ufa 450075, RB, Russia}
\adr b {Zel'dovich International Center for Astrophysics, M. Akmullah Bashkir State Pedagogical University, 3A, October Revolution Street, Ufa, 450077, RB, Russia}
\adr c {Department of Mathematics, Kidderpore College, 2, Pitamber Sircar Lane, Kolkata, 700023, WB, India}
}


\Abstract
	{The Damour-Solodukhin wormhole (hereinafter DSWH) is known to mimic Schwarzschild black hole (hereinafter SBH) horizon in some properties. To act as a mimicker, the DSWH parameter $\lambda$ by definition is required to be extremely tiny, i.e., $\lambda \sim 0$. Our comparative analyses show that such a requirement may be too restrictive at least as far as the Bondi accretion profiles of the two objects, the DSWH and SBH, are concerned. Intriguingly, it turns out that some profiles of DSWH mimic those for the SBH near the horizon even at values of $\lambda$ considerably higher, i.e., $\lambda \sim 1$.}
\medskip

] 
\email 1 {yu.rose@mail.ru\\ \cm (Corresponding author)}
\email 2 {karimov\_ramis\_92@mail.ru}
\email 3 {bamrita323@gmail.com}

\section{Introduction}

The spherically symmetric Bondi model has become a fundamental model in the study of accretion onto various astrophysical objects. The first studies of the gas-dynamic flow of substances begin with the seminal works of Hoyle and Littleton \cite{Hoyle:1939}, Bondi and Hoyle \cite{Bondi:1944}, in which the non-relativistic flow was considered. Later, Bondi considered accretion onto a star in the Newtonian approximation \cite{Bondi:1952}. Next, Michel \cite{Michel:1972} generalized the theory for a relativistic object by considering the accretion to a SBH, so it should more correctly be called the Bondi-Michel accretion process, but here we use the more popular term, i.e., Bondi accretion. To date, a lot of work has been done to study Bondi accretion onto various astrophysical objects. The main results can be found, e.g., in \cite{Yusupova:2021, Babichev:2005, Babichev:2013, Bahamonde:2015, Debnath:2015a, Debnath:2015b, Zeldovich:1971, Martnez-Pas:2014, Mach:2013, Karkowski:2006, Shatskiy:2010, Shatskii:1999, Ditta:2020, Debnath:2020a, Babichev:2011, Debnath:2020b, Donmez:2024, Pepe:2012, Babichev:2008, Dokuchaev:2011, Mitra:2024, Prust:2024}. It is important to note that accretion onto a central compact object considered here is self-consistent if: 1) mass of central compact objects changing slowly (quasi-static state, i.e. $\dot{M}(r)/M<<c_{s}/M$) and 2) the accreting matter is light, i.e. $M_{fluid}/M<<1$ \cite{Babichev:2013}.

Over the past decades, with the development of numerical modeling, one could obtain quite interesting results and visualize the accretion process. For example, a recent paper \cite{Combi:2024} presented the first dynamic model of plasma accretion to a static Simpson-Visser wormhole. The authors showed that the accreting plasma on one side of the throat settles in the throat and forms a hot rotating cloud, which subsequently dissipates from other side as relativistic heat flow.

The original DSWH \cite{Damour:2007} metric differs from the SBH metric in $g_{00}$ component designated by a dimensionless parameter $\lambda$. The wormhole with two flat regions has an antiphoton sphere for $\lambda <1/\sqrt{2}$ and a photon sphere for $\lambda \geq $ $1/\sqrt{2}$. When $\lambda =1/\sqrt{2}$, the spheres coincide and becomes marginally unstable because of this degeneracy \cite{Tsukomoto:2020}. In any case, stability issues are not the object of the present paper - we assume without loss of generality that the non-degenerate DSWH is stable.

It is known that the DSWH mimics many properties of a SBH \cite{Karimov:2019, Nandi:2022, Ovgun:2018, Sokoliuk:2022, Nandi:2018}. The physical reason why DSWH is dubbed as a black hole mimicker is the following: The first order coordinate time interval $\Delta t = 2M \log\left(\frac{1}{\lambda^{2}}\right)$ \cite{Damour:2007} taken by light to connect the throat to asymptotic region could be very large for a tiny $\lambda$ so that the two objects, the wormhole and the black hole, would be indistinguishable within the limited time scale of observation. However, Lemos and Zaslavskii \cite{Lemos:2008} argue that DSWH are bad mimickers of black holes, one of the reasons being the effect of infinitely large tidal effects on the freely falling body at the throat in the limit $\lambda \rightarrow 0$. This divergence is in stark contrast to the finite tidal effects occuring on a black hole horizon and this contrast can betray the purported mimicry. It should however be emphasized that, to avoid such divergent tidal effects on the throat, the original key assumption in \cite{Tsukomoto:2020} was \textit{not} to take the continuous limit $\lambda \rightarrow 0$ but consider \textit{only discrete small values of} $\lambda$ to be used in the first order of coordinate time interval $\Delta t$. This means that, for extremely tiny non-zero values of $\lambda $, say of the order of $0<\lambda <<\exp \left( -10^{15}\right) $ \cite{Damour:2007}, the tidal forces could be very large but  \textit{not} infinite.

Under the above circumstances, it would be of interest to study how the parameter $\lambda$ affects the Bondi accretion profiles of baryonic and phantom matter. We shall consider the profiles of radial velocity, density and the accretion rate in a steady-state accretion process of perfect fluids with barotropic equation of state $p=\omega \rho$ and compare them with those of the SBH ($\lambda = 0$). We shall consider cosmologically relevant accreting matter, i.e., phantom, quintessence, dust and stiff matter. We shall assume units in which $8\pi G=1$, $c=1$, unless specifically restored.

\section{Accretion onto Damour - Solodukhin wormhole}

Damour and Solodukhin proposed a simple but curious type of wormhole spacetime, also called a black hole foil, described by the metric in standard coordinates as
\bearr
	d\tau ^{2} = - A(r) dt^{2} + B(r)^{-1} dr^{2} + C(r) d\theta^{2}
\nnn
	+ C(r) \sin^{2}\theta d\varphi^{2},
\ear
with
\bearr
	A(r) =1-\frac{2M}{r}+\lambda ^{2}, \quad B(r)=1-\frac{2M}{r},
\nnn
	C(r)=r^{2}.
\ear

The throat of the wormhole is located at $r_{th} = 2M.$ When $\lambda =0$, one recovers SBH metric with an event horizon located at the radius $r = 2M$. By contrast, when $\lambda \neq 0$ the structure of the spacetime is different -- there is no event horizon, instead there is a throat that joins two isometric, asymptotically flat regions. This spacetime is an example of a Lorentzian wormhole \cite{Tsukomoto:2020}.

To study Bondi accretion to DSWH, we follow the procedure of Bahamonde and Jamil \cite{Bahamonde:2015}, who treated the accreting fluid particles as test particles neglecting their gravitational field. Fluid particles could be interacting among themselves so that pressure $p\neq 0$ except in the case of dust when $p=0$. The steadiness of the flow assumes that the mass increases slowly such that the distribution of the fluid has time to adjust itself to the changing metric. The treatment in \cite{Bahamonde:2015} was focused on black holes, whereas we shall apply the method here to wormholes.

Assuming the energy-momentum tensor of a perfect fluid, viz.,
\beq
	T_{\mu \upsilon} = (\rho +p) u_{\mu} u_{\upsilon} + p g_{\mu \upsilon},
\eeq
where $\rho$ is the energy density, $p$ the isotropic pressure and $u_{\mu}$ is the 4-velocity, which in general is
\beq
	u^{\mu} = \frac{dx^{\mu}}{d\tau} = \left(u^{t}, u^{r}, 0, 0\right),
\eeq
where $\tau$ is the proper time. We have $u^{\theta} = 0$ and $u^{\phi} = 0$ due to the assumption of radial motion of the accreting fluid. Note that all components of 4 -velocity, pressure, and energy density are functions of $r$ only. Since the 4-velocity must satisfy the normalization condition $u_{\mu} u^{\mu} = -1$, we find
\beq
	u^{t} := \frac{dt}{d\tau} = \sqrt{\frac{r\left\{r \left(1 + u^{2}\right) - 2M\right\}}{\delta 			\left(r - 2M\right)}},
\eeq
where for simplicity we have named $u = u^{r} = \frac{dr}{d\tau}$ and $\delta = r \left(1 + \lambda^{2}\right) - 2M$. Due to the presence of square root, there are two possibilities: $u>0$ $(<0)$, which respectively imply outward flow away from the accreting wormhole ($r$ increasing, so $u>0$) and inward flow toward the accreting wormhole ($r$ decreasing, so $u<0$). In the astrophysical context, both inward and outward flows are important. The former leads to change of the wormhole mass, while the latter leads to jets. Using the energy--momentum conservation law defined by 
\beq
	0 = T_{;\mu}^{\nu\mu} = \frac{1}{\sqrt{-g}}(\sqrt{-g} T^{\nu\mu})_{,\mu} + 			      \Gamma_{\alpha{\mu}}^{\nu} T^{\alpha\mu},
\eeq
we find after integrating
\beq
	\frac{\left(\rho + p\right) u r^{3/2} \delta \sqrt{r \left(1 + u^{2}\right) - 2M}}{r - 2M} 
	= A_{1},
\eeq
where $A_{1}$ is an integration constant.

\begin{figure*}
\centering
\includegraphics[width=14cm]{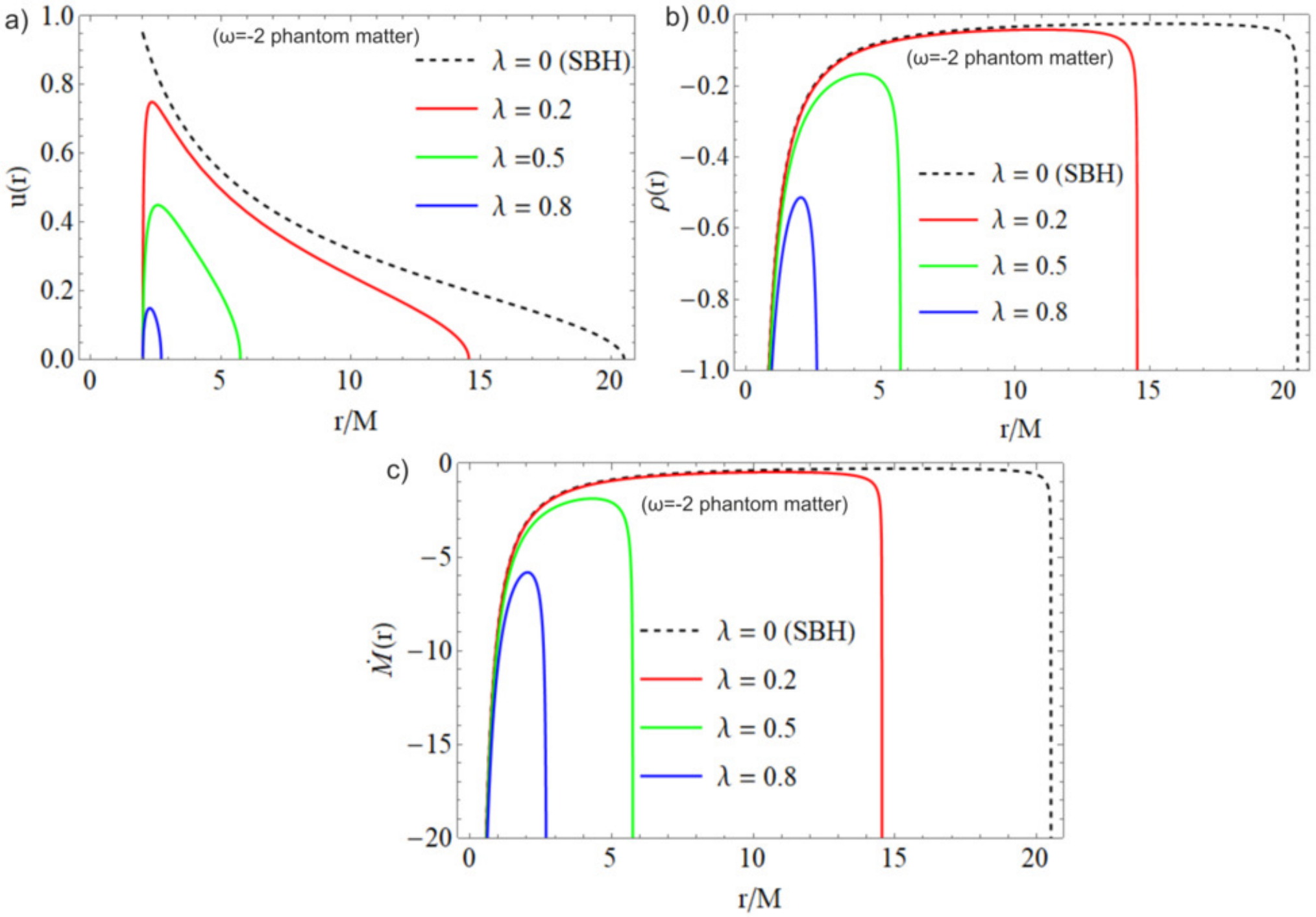}
\caption{\small 
		Velocity profile (a), energy density (b) of phantom energy ($\omega = -2$) and rate of change of mass (c) of DSWH and SBH (dashed line) versus $\frac{r}{M}$ for different values of $\lambda$. For illustration, we used the set of constants $A_{2} = 1$ and $A_{4} = 0.95$.}
		\label{fig1}
\end{figure*} 

\begin{figure*}
\centering
\includegraphics[width=14cm]{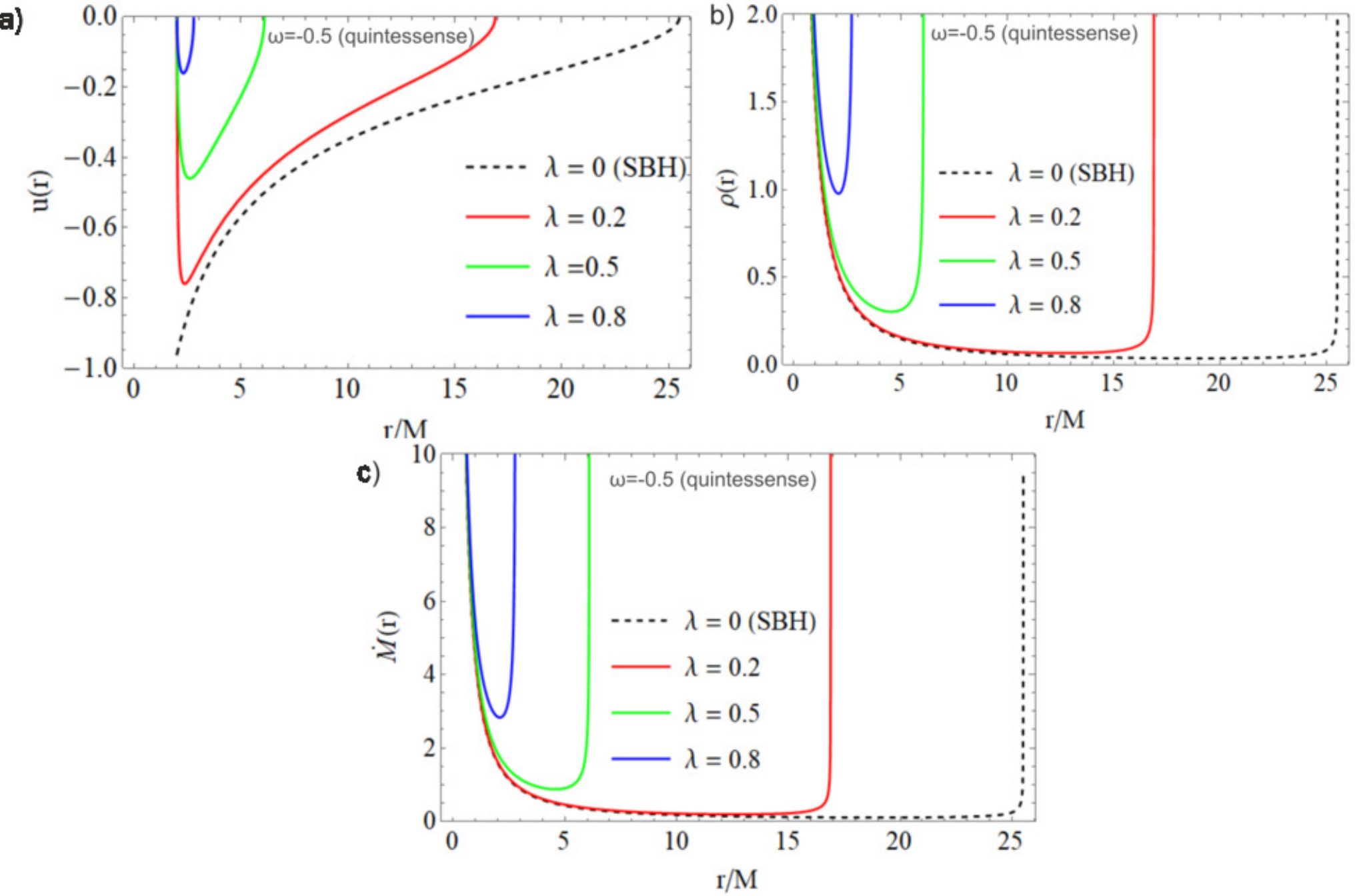}
\caption{\small 
		Velocity profile (a), energy density (b) of quintessence ($\omega = -0.5$) and rate of change of masses (c) of DSWH and SBH (dashed line) versus $\frac{r}{M}$ for different values of $\lambda$. For illustration, we used the set of constants $A_{2} = 1$ and $A_{4} = 0.48$.}  
		\label{fig2}
\end{figure*} 

By projecting the conservation law onto the 4-velocity $u_{{\mu}} T_{;\nu}^{\mu\nu} = 0$ and contracting all indices, we can find the relativistic energy flux (or continuity) equation
\beq
	u^{\mu}\rho_{,\mu} + (\rho +p)u_{;\mu}^{\mu}=0.
\eeq
We will assume that the pressure and energy density are related by a certain equation of state $p = p(\rho)$. The Eq.(8) for the generic metric (1) works out to
\beq
	\frac{\rho^{\prime}}{\rho + p} + \frac{u^{\prime}}{u} + \frac{A}{2A} + \frac{B^{\prime}}		{2B} + \frac{C^{\prime}}{C} = 0,
\eeq
where prime denotes differentiations with respect to $r$.

Integration of Eq.(9) yields
\beq
	\left(\rho + p\right) u r^{2} \sqrt{\frac{\delta}{r - 2M}} = - A_{0},
\eeq
where $A_{0}$ is an integration constant, while a negative sign is introduced on the right hand side since $u < 0$ on the left hand side is needed for accretion. Now, if we combine the above equation with (7), we obtain
\beq
	\sqrt{1 - \frac{2M}{r} + u^{2}}\sqrt{\frac{\delta}{r - 2M}} = A_{3},
\eeq
where $A_{3}$ is another constant. Futhemore, the equation of mass flux
\beq
	0 = J_{;\mu}^{\mu} = \frac{1}{\sqrt{-g}} \frac{d}{dr}(J^{r}\sqrt{-g}),
\eeq
leads to
\beq
	\rho u r^{2} \sqrt{\frac{\delta}{r - 2M}} = A_{2},
\eeq
where $A_{2}$ is an integration constant. If we divide Eq. (7) by (13), we get another useful relation,
\beq
	\frac{p + \rho}{\rho}\sqrt{1 - \frac{2M}{r} + u^{2}} \sqrt{\frac{\delta}{r - 2M}} = A_{4},
\eeq
where $A_{4}$ is yet another arbitrary constant. It should be noted here that integration constant $A_{1}$ from Eq. 7 can be represented as $A_{1} = A_{2} A_{4}$.

Taking differentials of Eqs. (13) and (14) and solving together, we obtain
\bearr
	\left[V^{2} - u^{2} \left(1 - \frac{2M}{r} + u^{2}\right) \right] \frac{du}{u} 
	+ \left(V^{2} - 1\right) 
\nnn
	\times \left[ \frac{2 \left(\delta + M\right)}{r \delta} - \frac{2M}{r \left(r - 2M\right)} \right.
\nnn
	\left. -\frac{M}{r \left\{ r \left(1 + u^{2}\right) - 2M\right\} } \right] dr = 0,
\ear 
where we have intoduced the parameter with the dimenison of velocity
\beq
	V^{2}=\frac{d\ln \left( \rho +p\right) }{d\ln \rho }-1.
\eeq

By taking the two brackets in (15) equal to zero, we can find the critical point of accretion located at $r = r_{c}$. Sonic points or critical points are the points at which the velocity of the moving fluid definitely equals the sound speed and maximum accretion rate occurs at the sonic point. Thus, we can decouple the quantities $u_{c}^{2}$ and $V_{c}^{2}$ and obtain
\beq
	u_{c}^{2}=\frac{M\left(r_{c} - 2M\right)}{2 r_{c} \delta_{c}},
\eeq
\beq
	V_{c}^{2}=\frac{M}{2\delta_{c} + M}.
\eeq
Here every function is evaluated at $r=r_{c}$, $\delta_{c} = \delta |_{r=r_{c}}$ and $u_{c}$ is the critical speed of the flow. The speed of sound is found to be
\beq
	c_{s}^{2} = \frac{\partial p}{\partial \rho} = A_{4} \sqrt{\frac{r_{c}\left(r_{c} - 2M\right)}{\delta_{c}\left\{r_{c}\left(1 + u_{c}^{2}\right) - 2M\right\}}}-1.
\eeq

The rate of change of mass is given by \cite{Debnath:2015a}
\beq
	\dot{M}=4\pi A_{3}M^{2}(\rho +p),
\eeq
where the dot represents derivative with respect to time and $A_{3} = A_{2}A_{4}/(1 + \omega)$ as in \cite{Bahamonde:2015}.

It is possible to integrate the conservation laws and obtain analytical expressions of the physical parameters. Here, for simplicity, we study the accreting barotropic fluid with an equation of state $p = \omega \rho$. Using Eqs. (7) and (14), we can find directly one set of solutions for radial velocity, density and the rate of change of the mass
\beq
	u(r) = - \frac{\sqrt{r - 2M} \sqrt{A_{4}^{2}r - \delta \left(1 + \omega\right)^{2}}}{\left(1 + 		\omega\right)\sqrt{\delta r}},
\eeq
\beq
	\rho(r) = \frac{A_{2} \left(1 + \omega\right)}{r^{3/2} \sqrt{A_{4}^{2}r - \delta
	\left(1 + \omega\right)^{2}}},
\eeq
\beq
	\dot{M}(r) = \frac{4 A_{2}^{2}A_{4}M^{2}\pi \left(1 + \omega\right)}{r^{3/2} 
	\sqrt{A_{4}^{2}r - \delta \left(1 + 	\omega\right)^{2}}}.
\eeq

\begin{figure*}
\centering
\includegraphics[width=14cm]{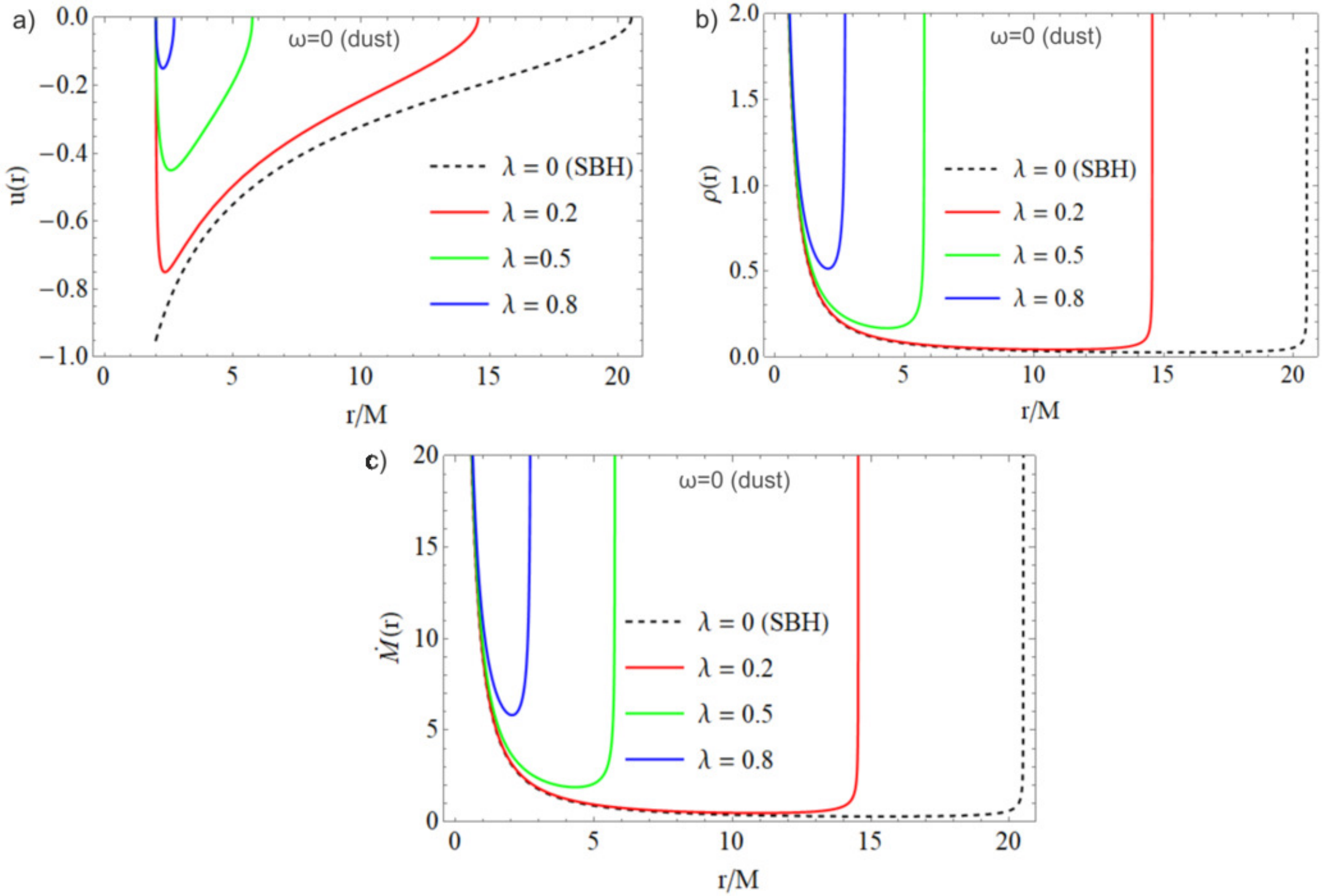}
\caption{\small 
		Velocity profile (a), energy density (b) of dust ($\omega = 0$) and rate of change of masses (c) of DSWH and SBH (dashed line) versus $\frac{r}{M}$ for different values of $\lambda$. For illustration, we used the set of constants $A_{2} = 1$ and $A_{4} = 0.95$.}
		\label{fig3}
\end{figure*} 

\begin{figure*}
\centering
\includegraphics[width=14cm]{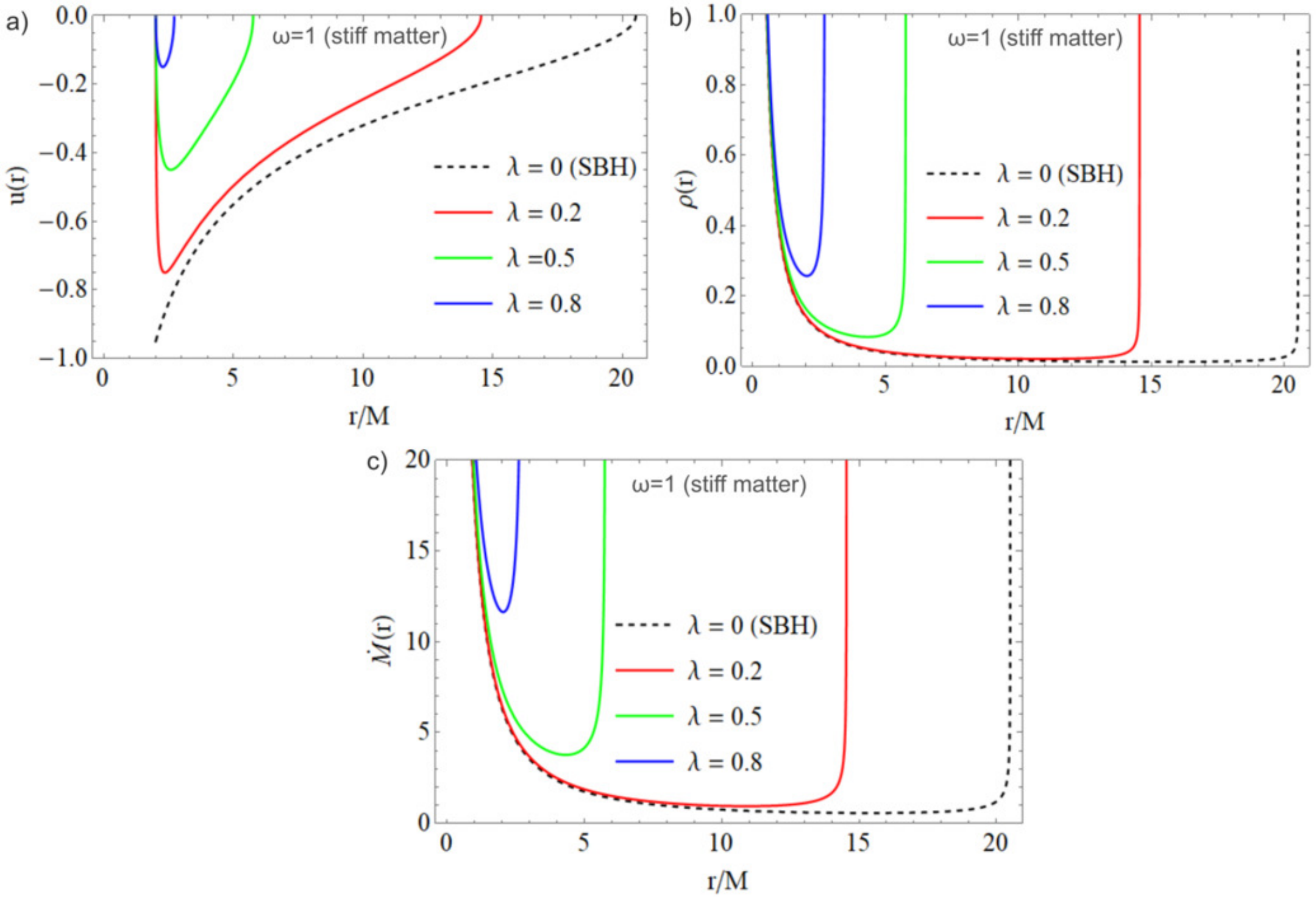}
\caption{\small 
		Velocity profile (a), energy density (b) of stiff matter ($\omega = 0$) and rate of change of masses (c) of DSWH and SBH (dashed line) versus $\frac{r}{M}$ for different values of $\lambda$. For illustration, we used the set of constants $A_{2} = 1$ and $A_{4} = 1.9$.}
		\label{fig4}
\end{figure*} 

Combining Eqs. (17) and (21), we get two expressions for critical radius $r_{c}$
\beq
	r_{c} = 2M,
\eeq
\beq
	r_{c}=\frac{3M(1+\omega )^{2}}{2((1+\omega )^{2}(1+\lambda ^{2})-A_{4}^{2})}.
\eeq
 
We consider only second root (Eq.25), because it depends on $\omega$ and $\lambda$. The critical point must be positive, i.e. from Eq. (25) we require
\beq
	(1 + \omega)^{2} (1 + \lambda^{2}) - A_{4}^{2} > 0
\eeq
and
\beq
	\frac{3M(1 + \omega)^{2}}{2((1 + \omega)^{2}(1 + \lambda^{2}) - A_{4}^{2})} > r_{th}.
\eeq

Also, the last two conditions allow to set the upper (Eq.26) and lower (Eq.27) limit of the value for $A_{4}$, which are shown in the Table 1. The $A_{4}>0$ is determined by the positive real values of the radical in Eqs. (26) and (27).

\begin{table}
\caption{Intervals of $A_{4}$ for different values of parameter of state $\omega$ and parameter $\lambda $.}
\centering
\footnotesize
\begin{tabular}{|c|c|c|}
\hline
$\omega $ & $\lambda $ & Intervals of $A_{4}$ from Eqs.(27,28) \\ \hline
$-2$ & $0$ & $0.5<A_{4}<1$ \\ 
- & $0.2$ & $0.5385<A_{4}<1.0198$ \\ 
- & $0.5$ & $0.7071<A_{4}<1.1180$ \\ 
- & $0.8$ & $0.9434<A_{4}<1.2806$ \\ 
- & $1$ & $1.1180<A_{4}<1.4142$ \\
\hline
$-0.5$ & $0$ & $0.25<A_{4}<0.5$ \\ 
- & $0.2$ & $0.2693<A_{4}<0.5099$ \\ 
- & $0.5$ & $0.3535<A_{4}<0.5590$ \\ 
- & $0.8$ & $0.4717<A_{4}<0.6403$ \\ 
- & $1$ & $0.5590<A_{4}<0.7071$ \\
\hline
$0$ & $0$ & $0.5<A_{4}<1$ \\ 
- & $0.2$ & $0.5385<A_{4}<1.0198$ \\ 
- & $0.5$ & $0.7071<A_{4}<1.1180$ \\ 
- & $0.8$ & $0.9434<A_{4}<1.2806$ \\ 
- & $1$ & $1.1180<A_{4}<1.4142$ \\
\hline
$1$ & $0$ & $1<A_{4}<2$ \\ 
- & $0.2$ & $1.0770<A_{4}<2.0396$ \\ 
- & $0.5$ & $1.4142<A_{4}<2.2361$ \\ 
- & $0.8$ & $1.8868<A_{4}<2.5612$ \\ 
- & $1$ & $2.2361<A_{4}<2.8284$ \\
\hline
\end{tabular}
\end{table}
\normalsize

Now we fix the sings and values of integration constants based on relevant
equations to study accretion process for different kinds of fluid. For the
accretion of quintessense, dust and phantom matter to take place, we need to
have $u(r)<0,\rho (r)>0$, but in the case of phantom matter $u(r)>0$, $\rho
(r)<0$ $\Longrightarrow A_{2}>0$ from Eq. (22). So, to plot radial velocity,
energy density and the rate of change of mass for different types of matters
we take $A_{2}=1$, and $\ $values of $A_{4}$ we can choose from the Table 1,
i.e. each parameter of state $\omega $ has its own specific value of $A_{4}$%
. The value of $A_{4}$ for each parameter of state $\omega $ must be such
that it enters all the relevant intervals from $\lambda =0$ to $\lambda =1$.
Plots of the accretion process profiles of matters onto SBH and DSWH are
shown in Figures 1-4.

\section{Results}

Figs. 1--4 represent radial velocity, energy density, and rate of change of
central mass profiles for different values of the state parameter $\omega $.
Here, $\omega =-2$ (Figure 1), $\omega =-0.5$ (Figure 2), $\omega =0$
(Figure 3), and $\omega =1$ (Figure 4) refer to phantom, quintessence, dust,
and stiff matter, respectively. We consider the central masses are
numerically the same ($M=1$).

Fig. 1a represents radial velocity profiles of phantom fluid ($\omega =-2$)
versus radial coordinate $r/M$ for DSWH and SBH. As figure shows, the DSWH
radial velocity profiles remain lower than that of SBH\ at all radii. In
addition, increasing the values of $\lambda $ also lowers the values of
radial velocity $u(r).$ In the case of SBH at the horizon radius $r_{h}=2M$
the radial velocity function $u(r)$ tends to a finite value, i.e. $u(r)\neq 0
$, but for DSWH at the throat radius $r_{th}=2M$ the radial velocity
function goes to zero $u(r)=0$.

Fig. 1b represents the behavior of the energy density function $\rho (r)$ of
phantom energy. It can be seen that the values of energy density of the
phantom fluid to DSWH is less than to SBH. In the case of DSWH and SBH,
absolute values of density increase in the vicinity of $r=2M$.

Fig. 1c represents the rate of change of mass against the radial coordinate $%
r/M$. As the plots show, the accretion of phantom energy deacreases the
masses of DSWH and SBH, since $\dot{M}(r)<0$. With the increase of
values of $\lambda$, the rate of change of DSWH mass increases.

For quintessence (Fig.2), dust (Fig.3) and stiff matter (Fig.4), Figs.
2a,3a,4a show that DSWH has the lower velocity of accreting fluids, and the
SBH profile shows higher velocity profiles. In the case of DSWH, the
increase in $\lambda$ induces a decrease in radial velocity of the
accreting fluids. In the case of SBH at the horizon $r_{h}=2M$ the radial
velocity function $u(r)$ tends to a finite value, but for DSWH at the throat 
$r_{th}=2M$ the radial velocity function $u(r)=0$ for all types of accreting
fluids.

Figs. 2b,3b,4b represents density profiles of the accreting fluid. It can be
seen that the highest density is achieved near the $r=2M$ (for DSWH and SBH)
and tend to bunch together. Density profiles of DSWH are higher than the
density profiles of SBH but coinside near $r=2M$.

Figs. 2c,3c,4c show that the non-phantom accretion increases masses of DSWH
and SBH. As we can see, the profiles of rate of change of mass $\dot{M}
(r)$ of non-phantom fluids are same in the vicinity of $r=2M$ for two
central objects. The rate of change of DSWH mass increasing with the
increase values of $\lambda$.

From Eqs.(21) and (22) we note that $u(r)$ diverges and $\rho (r) = 0$ at the "phantom divide $\omega = -1$". It means that dark energy with  $\omega = -1$ cannot accrete to DSWH. For example, the similar results were obtained in \cite{Yusupova:2021} where authors studied Bondi accretion onto Ellis-Bronnikov wormhole. This divergence reflects a pathology that is known to occur at the divide; viz., the fluid perturbations become divergent, meaning that the adiabatic sound speed $c_{s}^{2}$ diverges at the crossing $\omega = -1$, that was also demonstrated by Kunz and Sapone \cite{Kunz:2006}. Srivastava in \cite{Srivastava:2006} showed that the crossing the divide is possible if one considers the "scale-factor-dependent" equation of state $p=-\rho +f(a)$, where $f$ is a function of the scale factor $a$.

\section{Conclusions}

In the above, we studied the steady-state Bondi accretion process of perfect fluids onto DSWH, the latter reduceing to the SBH at $\lambda =0.$ We analyzed the profiles of radial velocity $u(r)$, the density $\rho (r)$ and the rate of change of mass $\dot{M}$ of the central gravitating source due to flow of barotropic fluids with equation of state $p=\omega \rho $, where $\omega $ describes the type of accreting fluid, and compare the profiles with those of the SBH. We considered cosmologically relevant accreting fluids, i.e., phantom ($\omega =-2$), quintessence ($\omega =-0.5$), dust ($\omega =0$) and stiff matter ($\omega =1$) and the results are indicated in the plots. We fixed the interval of the integration constant ($A_{4}$), as shown in Table 1, using the physical conditions that the critical radius $r=r_{c}$ must exceed the throat and the horizon. It is found that the accretion profiles of the considered fluids first decrease and then increase near the event horizon of SBH and the throat of DSWH, both occurring at $r=2M$. The relevant equations show that the dark energy ($\omega =-1$) cannot accrete due to the divergences appearing in the profiles.

The propagation of light from the throat of DSWH out to a distant observer can take logarithmically infinite coordinate time $\Delta t$ only at $\lambda \sim 0$ thereby mimicking the property of the SBH\ horizon. Such restriction of tiny values of $\lambda $ may be waived, when fluid accretion is concerned. Interestingly, as shown in Figs. 1-4, all the accretion profiles in the close vicinity of the SBH horizon and the DSWH throat are nearly the same, which show that the DSWH mimics SBH even at \textit{high} values of $\lambda \sim 1$ at least as far as Bondi accretion is concerned.

These results can be used for interpretation of the future observation data of Millimetron - type projects \cite{Kardashev:2014, Andrianov:2021}.

\Acknow{The authors thank two anonymous referees for their useful suggestions that
helped improve the paper.}

\Funding{This work was supported by the Russian Science Foundation under grant no.
23-22-00391
\\
https://rscf.ru/en/project/23-22-00391/.}


\end{document}